\documentstyle[floats,prl,aps]{revtex}
\begin{document}
\renewcommand{\thefootnote}{\fnsymbol{footnote}}
\draft
\title{\large\bf 
  Integrable boundary conditions for the $q$-deformed extended 
  Hubbard model}

\author{  Xiang-Yu Ge \footnote {E-mail:xg@maths.uq.edu.au}}
\address{Centre for Mathematical Physics,Department of Mathematics,
University of Queensland, Brisbane, Qld 4072, Australia}

\maketitle

\vspace{10pt}

\begin{abstract}
Integrable open-boundary condition for the $q$-deformed 
Essler-Korepin-Schoutens extended Hubbard model of strongly
correlated electrons, 
are studied in the framework of the boundary quantum inverse scattering method. 
Diagonal boundary $K$-matrices are found, and nine classes of integrable
boundary terms are determined.
\end{abstract}




\def\a{\alpha}
\def\b{\beta}
\def\d{\delta}
\def\e{\epsilon}
\def\g{\gamma}
\def\k{\kappa}
\def\l{\lambda}
\def\o{\omega}
\def\t{\theta}
\def\s{\sigma}
\def\D{\Delta}
\def\L{\Lambda}


\def\beq{\begin{equation}}
\def\eeq{\end{equation}}
\def\bea{\begin{eqnarray}}
\def\eea{\end{eqnarray}}
\def\ba{\begin{array}}
\def\ea{\end{array}}
\def\no{\nonumber}
\def\le{\langle}
\def\re{\rangle}
\def\lt{\left}
\def\rt{\right}

\newcommand{\reff}[1]{eq.~(\ref{#1})}

\vskip.3in

One-dimensional strongly correlated electron systems with boundaries are
of great importance because of their promising role in theoretical 
condensed matter
physics and possibly in high-$T_c$ superconductivity\cite{Ess92/3}.
Boundary conditions and nontrivial
boundary interactions for such systems, which are compatible with
integrability in the bulk, are constructed from solutions of the graded
reflection equations \cite{Skl88}, and have attracted much attention recently in
connection with physical problems like $X$-ray edge singularities
\cite{Ess97}, orthogonalities catastrophy \cite{VP98} and tunneling through
constrictions \cite{Aba97} in quantum wires. In particular, open
boundaries and boundary fields for the Hubbard-like models 
\cite{Suz96,Shi97,Yam97,Bed97}
and for the supersymmetric $t-J$ model \cite{Gon94,Ess96,Asa97} have been
studied in connection with this.
The results of the present letter may well have interesting
applications to these problems.

In this letter, we shall construct the open-boundary conditions for the
$q$-deformed Essler-Korepin-Schoutens extended Hubbard model (EKS model)
\cite{Ess92/3} which preserve the integrability of the model. This is achieved
by solving the graded reflection equations for the diagonal boundary
$K$-matrices.

Let $c_{j,\s}$ and $c_{j,\s}^{\dagger}$ denote fermionic creation and
annihilation operators for spin $\s$ at
site $j$, which satisfy the anti-commutation relations 
$\{c_{i,\s}^\dagger, c_{j,\tau}\}=\d_{ij}\d_{\s\tau}$, where 
$i,j=1,2,\cdots,L$ and $\s,\tau=\uparrow,\;\downarrow$. We consider 
the open-boundary $q$-deformed EKS model with  
Hamiltonian of the form:
\beq
H=\sum _{j=1}^{L-1} H_{j,j+1}^Q + H^{\rm boundary}_{lt} +H^{\rm
boundary}_{rt},\label{h}\label{hamiltonian-b}
\eeq
where $H^{\rm boundary}_{lt}~ (H^{\rm boundary}_{rt})$ stand for left 
(right) boundary terms whose explicit  
forms are spelled out below, and
$H_{j,j+1}^Q$ is the bulk Hamiltonian density of the $q$-deformed
EKS model: 
\bea
H_{j,j+1}^Q&=&\sum_\a(c_{j,\a}^\dagger c_{j+1,\a}+{\rm h.c.})
  (1-n_{j,-\a}-n_{j+1,-\a})
  \no\\
 & &+c^\dagger_{j\uparrow} c^\dagger_{j\downarrow}
 c_{j+1,\downarrow} c_{j+1,\uparrow}
 +c^\dagger_{j+1,\uparrow} c^\dagger_{j+1,\downarrow}
 c_{j\downarrow} c_{j\uparrow}
 -{S}_j ^+{S}_{j+1}^--S_j^-S_{j+1}^+\no\\
 & &- q^{-1}((n_{j\uparrow}
 - n_{j\downarrow})^2
 +n_{j+1,\uparrow} n_{j+1,\downarrow}
 -n_{j\uparrow} n_{j+1,\downarrow})\no\\
 & &- q((n_{j+1,\uparrow}
 - n_{j+1,\downarrow})^2
 +n_{j\uparrow} n_{j\downarrow}
 -n_{j+1,\uparrow} n_{j\downarrow}),
  \label{hamiltonian}
\eea
where $q$ is a free parameter, $n_{j}=n_{j,\uparrow}+n_{j,\downarrow}$ with
$n_{j,\s}=c_{j,\s}^\dagger c_{j,\s}$ being the density operator
for the fermion of species $\s$ at site $j$, and
$S^{+}$=$c_{\downarrow}^\dagger c_{\uparrow}$,$S^{-}$=$c_{\uparrow}^\dagger
c_{\downarrow}$. We omit the details of calculating
the above Hamiltonian,  but remark that the four states of the vector irrep
of the quantum algebra underlying the model,  $U_q[gl(2|2)]$,  are identified 
with the electronic states
\beq
|0\re\,,~~~|\uparrow\re=c_{\uparrow}^\dagger |0\re\,,~~~
|\downarrow\re=c_{\downarrow}^\dagger |0\re\,,
~~~|\uparrow,\downarrow\re=c_{\downarrow}^\dagger c_{\uparrow}^\dagger |0\re\,.  \label{states}
\eeq
The model defined by (\ref{hamiltonian}) is $U_q[gl(2|2)]$ supersymmetric
and is exactly solvable on the one-dimensional periodic lattice.
This is because the local Hamiltonian $H_{j,j+1}^Q$
is actually derived through the QISM
using a $U_q[gl(2|2)]$ invariant $R$-matrix. 
To show this, we denote the generators of $U_q[gl(2|2)]$ by $E^\mu_\nu,~~
\mu,\nu=1,2,3,4$ with grading $[|1\re]=[|2\re]=0,~[|3\re]=[|4\re]=1$.
In a typical 4-dimensional representation $V(\L)$ of $U_q[gl(2|2)]$, 
the highest weight is $\L=(1,0|0,0)$ .
Let $\{|x\re\}_{x=1}^4$
denote an orthonormal basis with $|1\re, |2\re$ even
(bosonic) and $|3\re, |4\re$ odd (fermionic). Then
the simple generators $\{E^i_j\}=e_{ij}(i,j=1,2,3,4)$
are $4\times 4$ supermatrices.
Associated with $U_q(gl(2|2))$ there is a graded  
coproduct structure $\D:U_q[gl(2|2)]\rightarrow U_q[gl(2|2)]\otimes
U_q[gl(2|2)]$ given by
\bea
\D(E^\mu_\mu)&=&I\otimes E^\mu_\mu+E^\mu_\mu \otimes I,~~~~ \mu=1,2,3,4,\no\\
\D(E^1_{2})&=&E^1_{2} \otimes q^{\frac{1}{2}(E^1_1-
   E^{2}_{2})}+ q^{-\frac{1}{2}(E^1_1-E^{2}_{2})} \otimes E^1_{2},~~~
\D(E^{2}_{1})=E^{2}_{1} \otimes q^{\frac{1}{2}(E^1_1- E^{2}_{2})}+
   q^{-\frac{1}{2}(E^1_1-E^{2}_{2})} \otimes E^{2}_{1},\no\\
\D(E^2_{3})&=&E^2_{3} \otimes q^{\frac{1}{2}(E^2_2+
   E^{3}_{3})}+ q^{-\frac{1}{2}(E^2_2+E^{3}_{3})} \otimes E^2_{3},~~~
\D(E^{3}_{2})=E^{3}_{2} \otimes q^{\frac{1}{2}(E^2_2+ E^{3}_{3})}+
   q^{-\frac{1}{2}(E^2_2+E^{3}_{3})} \otimes E^{3}_{2},\no\\
\D(E^3_{4})&=&E^3_{4} \otimes q^{\frac{1}{2}(E^4_4-
   E^{3}_{3})}+ q^{-\frac{1}{2}(E^4_4-E^{3}_{3})} \otimes E^3_{4},~~~
\D(E^{4}_{3})=E^{4}_{3} \otimes q^{\frac{1}{2}(E^4_4- E^{3}_{3})}+
   q^{-\frac{1}{2}(E^4_4-E^{3}_{3})} \otimes E^{4}_{3}.
\eea

Under the coproduct action the graded tensor product $V\otimes V$ is also a
$U_q[gl(2|2)]$ module which reduces completely: $V(\L)\otimes V(\L)=
V(\L_1)\oplus V(\L_2)$, where $V(\L_1)$ and $V(\L_2)$ are 8-dimensional
modules.  Associated with each 8-dimensional representation, there is 
a $U_q[gl(2|2)]$-invariant $R$-matrix
which satisfies the graded Yang-Baxter equation. The $R$-matrix
is given by
\beq
\check{R}(u)=\check{P}_1+\frac{1-q^{u+2}}{q^u-q^2}\check{P}_2,
 \label{rational-R}
\eeq
where $\check{P}_a:~ \check{P}_a[V(\L)\otimes V(\L)]=V(\L_a),~~a=1,2$,
are two projection operators which may be constructed as
\beq
\check{P}_1=\sum_{k=1}^8 |\Psi_k^1\re\le\Psi_k^1|,~~~~~~
  \check{P}_2=\sum_{k=1}^8|\Psi_k^2\re\le\Psi_k^2|.\label{projector1}
\eeq
Here$ |\Psi_k^a\re,~a=1,2,~ k=1,2,\cdots,8$ are basis vectors for
$V(\L_a)$.  Throughout this letter,
\bea
&&\le\Psi^a_k|=\left (|\Psi^a_k\re\right )^\dagger,~~~~
  \left (|x\re\otimes |y\re\right )^\dagger=(-1)^{[|x\re][|y\re] }
  \le y|\otimes \le x|\label{dual1}
\eea
with $[|x\re]=0$ for even (bosonic) $|x\re$, and $[|x\re]=1$ for odd
(fermionic) $|x\re$. The basis vectors $|\Psi^1_k\re,~ 
|\Psi_k^2\re,~k=1,2,\cdots,8$ are chosen as
\bea
|\Psi^1_1\re&=&|1\re\otimes |1\re,~~~
|\Psi^1_2\re=|2\re\otimes |2\re,\no\\
|\Psi^1_i\re&=&\frac{1}{\sqrt{q+q^{-1}}}(q^{\frac{1}{2}}
  |i-1\re\otimes |1\re+q^{-\frac{1}{2}}|1\re\otimes |i-1\re),~~~i=3,4,5,\no\\
|\Psi^1_i\re&=&\frac{1}{\sqrt{q+q^{-1}}}(q^{\frac{1}{2}}
  |i-3\re\otimes |2\re+q^{-\frac{1}{2}}|2\re\otimes |i-3\re),~~~i=6,7,\no\\
|\Psi^1_8\re&=&\frac{1}{\sqrt{q+q^{-1}}}(q^{\frac{1}{2}}
  |4\re\otimes |3\re-q^{-\frac{1}{2}}|3\re\otimes |4\re),\no\\
|\Psi^2_1\re&=&\frac{1}{\sqrt{q+q^{-1}}}(q^{\frac{1}{2}}
  |3\re\otimes |4\re+q^{-\frac{1}{2}}|4\re\otimes |3\re),\no\\
|\Psi^2_i\re&=&\frac{1}{\sqrt{q+q^{-1}}}(q^{\frac{1}{2}}
  |1\re\otimes |i\re-q^{-\frac{1}{2}}|i\re\otimes |1\re),~~~i=2,3,4,\no\\
|\Psi^2_i\re&=&\frac{1}{\sqrt{q+q^{-1}}}(q^{\frac{1}{2}}
  |2\re\otimes |i-2\re-q^{-\frac{1}{2}}|i-2\re\otimes |2\re),~~~i=5,6,\no\\
|\Psi^2_7\re&=&|3\re\otimes |3\re,~~~
|\Psi^2_8\re=|4\re\otimes |4\re.\label{basis1}
\eea

On the $L$-fold tensor product space $V\otimes V\otimes \cdots\otimes V$
we denote $\check{R}(u)_{j,j+1}=
I^{\otimes(j-1)}\otimes\check{R}(u)\otimes I^{\otimes(L-j-1)}$, and
define the local Hamiltonian by
\beq
H^{\rm R}_{j,j+1}=\left .\frac{d}{du}\check{R}_{j,j+1}
   (u)\right |_{u=0}
\eeq
We make the identifications:
\bea
&&|1\re=|0\re\,,~~~
  |2\re=c_{j,\downarrow}^\dagger c_{j,\uparrow}^\dagger|0\re\,,~~~ 
  |3\re=c_{j,\uparrow}^\dagger|0\re\,,~~~
  |4\re=c_{j,\downarrow}^\dagger|0\re\
  .\label{choice}
\eea
Then by (\ref{projector1}), (\ref{basis1})
(\ref{dual1}) and (\ref{choice}),
and after tedious but straightforward manipulation,
we get, up to an additive constant, 
\beq
H_{j,j+1}^Q=-\frac{q-q^{-1}}{\ln{q}}
   \;H^{\rm R}_{j,j+1}. 
\eeq
This identity also shows that the bulk part of $H$ with $H_{j,j+1}^Q$ as
in (\ref{hamiltonian}), 
commutes with the generators ${E^i_j}=e_{ij} (i,j=1,2,3,4)$ of $U_q[gl(2|2)]$, 
since the $R$-matrix $\check{R}(u)$ is $U_q[gl(2|2)]$ invariant. 

Now we propose the following nine classes of boundary conditions:
\bea
{\rm Case~ (i)}:~~ &&H^{\rm boundary}_{lt}=2\sinh \g
  \lt((1-\frac {e^{-\frac{\xi^I_-}{2}\g}}{2\sinh\frac{\xi^I_-}{2}\g})
  n_{1\uparrow}n_{1\downarrow} -n_1\rt),\no\\
&&H^{\rm boundary}_{rt}=-2\sinh\g
  \lt((1-\frac {e^{-\frac{\xi^I_+}{2}\g}}{2\sinh\frac{\xi^I_+}{2}\g})
  n_{L\uparrow}n_{L\downarrow} -n_L\rt)
  ;\label{boundary11}\\
{\rm Case~ (ii)}:~~ &&H^{\rm boundary}_{lt}=-2\sinh\g
  \lt(\frac {e^{\frac{\xi^{II}_-}{2}\g}}{2\sinh\frac{\xi^{II}_-}{2}\g}
  n_{1\uparrow}+n_{1\downarrow}
 - n_{1\uparrow}n_{1\downarrow} \rt),\no\\
&&H^{\rm boundary}_{rt}=2\sinh\g
  \lt(\frac {e^{\frac{\xi^{II}_+}{2}\g}}{2\sinh\frac{\xi^{II}_+}{2}\g}
  n_{L\uparrow}+n_{L\downarrow}
 - n_{L\uparrow}n_{L\downarrow} \rt)
  ;\label{boundary22}\\
{\rm Case~ (iii)}:~~ &&H^{\rm boundary}_{lt}=-\frac{e^{\frac{\xi^{III}_-}{2}\g}
  \sinh \g}{\sinh\frac{\g\xi^{III}_-}{2}}
  \lt( n_{1\uparrow}+n_{1\downarrow}
  -2 n_{1\uparrow}n_{1\downarrow} \rt),\no\\
&&H^{\rm boundary}_{rt}=-\frac{e^{\frac{\xi^{III}_+}{2}\g}
  \sinh \g}{\sinh\frac{\g\xi^{III}_+}{2}}
  \lt( n_{L\uparrow}+n_{L\downarrow}
  -2 n_{L\uparrow}n_{L\downarrow} \rt)
  ;\label{boundary33}\\
{\rm Case~ (iv)}:~~ &&H^{\rm boundary}_{lt}=2\sinh \g
  \lt((1-\frac {e^{-\frac{\xi^I_-}{2}\g}}{2\sinh\frac{\xi^I_-}{2}\g})
  n_{1\uparrow}n_{1\downarrow} -n_1\rt),\no\\
&&H^{\rm boundary}_{rt}=2\sinh\g
  \lt(\frac {e^{\frac{\xi^{II}_+}{2}\g}}{2\sinh\frac{\xi^{II}_+}{2}\g}
  n_{L\uparrow}+n_{L\downarrow}
 - n_{L\uparrow}n_{L\downarrow} \rt)
  ;\label{boundary12}\\
{\rm Case~ (v)}:~~ &&H^{\rm boundary}_{lt}=-2\sinh\g
  \lt(\frac {e^{\frac{\xi^{II}_-}{2}\g}}{2\sinh\frac{\xi^{II}_-}{2}\g}
  n_{1\uparrow}+n_{1\downarrow}
 - n_{1\uparrow}n_{1\downarrow} \rt),\no\\
&&H^{\rm boundary}_{rt}=-2\sinh\g
  \lt((1-\frac {e^{-\frac{\xi^I_+}{2}\g}}{2\sinh\frac{\xi^I_+}{2}\g})
  n_{L\uparrow}n_{L\downarrow} -n_L\rt)
  ;\label{boundary21}\\
{\rm Case~ (vi)}:~~ &&H^{\rm boundary}_{lt}=2\sinh \g
  \lt((1-\frac {e^{-\frac{\xi^I_-}{2}\g}}{2\sinh\frac{\xi^I_-}{2}\g})
  n_{1\uparrow}n_{1\downarrow} -n_1\rt),\no\\
&&H^{\rm boundary}_{rt}=-\frac{e^{\frac{\xi^{III}_+}{2}\g}
  \sinh \g}{\sinh\frac{\g\xi^{III}_+}{2}}
  \lt( n_{L\uparrow}+n_{L\downarrow}
  -2 n_{L\uparrow}n_{L\downarrow} \rt)
  ;\label{boundary13}\\
{\rm Case~ (vii)}:~~ &&H^{\rm boundary}_{lt}=-\frac{e^{\frac{\xi^{III}_-}{2}\g}
  \sinh \g}{\sinh\frac{\g\xi^{III}_-}{2}}
  \lt( n_{1\uparrow}+n_{1\downarrow}
  -2 n_{1\uparrow}n_{1\downarrow} \rt),\no\\
&&H^{\rm boundary}_{rt}=-2\sinh\g
  \lt((1-\frac {e^{-\frac{\xi^I_+}{2}\g}}{2\sinh\frac{\xi^I_+}{2}\g})
  n_{L\uparrow}n_{L\downarrow} -n_L\rt)
  ;\label{boundary31}\\
{\rm Case~ (viii)}:~~ &&H^{\rm boundary}_{lt}=-2\sinh\g
  \lt(\frac {e^{\frac{\xi^{II}_-}{2}\g}}{2\sinh\frac{\xi^{II}_-}{2}\g}
  n_{1\uparrow}+n_{1\downarrow}
 - n_{1\uparrow}n_{1\downarrow} \rt),\no\\
&&H^{\rm boundary}_{rt}=-\frac{e^{\frac{\xi^{III}_+}{2}\g}
  \sinh \g}{\sinh\frac{\g\xi^{III}_+}{2}}
  \lt( n_{L\uparrow}+n_{L\downarrow}
  -2 n_{L\uparrow}n_{L\downarrow} \rt)
  ;\label{boundary23}\\
{\rm Case~ (ix)}:~~ &&H^{\rm boundary}_{lt}=-\frac{e^{\frac{\xi^{III}_-}{2}\g}
  \sinh \g}{\sinh\frac{\g\xi^{III}_-}{2}}
  \lt( n_{1\uparrow}+n_{1\downarrow}
  -2 n_{1\uparrow}n_{1\downarrow} \rt),\no\\
&&H^{\rm boundary}_{rt}=2\sinh\g
  \lt(\frac {e^{\frac{\xi^{II}_+}{2}\g}}{2\sinh\frac{\xi^{II}_+}{2}\g}
  n_{L\uparrow}+n_{L\downarrow}
 - n_{L\uparrow}n_{L\downarrow} \rt)
  .\label{boundary32}
\eea
where $\xi^I_{\pm}, \xi^{II}_{\pm}, \xi^{III}_{\pm} $ are parameters describing 
boundary effects. As will be shown below, all nine classes
of boundary conditions lead to integrable models.

Quantum integrability of the system defined
by Hamiltonian (\ref{hamiltonian}) with any of the nine boundary conditions
(\ref{boundary11}--{\ref{boundary32})
can be established as follows by means of the supersymmetric boundary QISM 
\cite{Skl88,Mez91,Veg93,Zho53,Bra98}. 
We first search for boundary $K$-matrices which satisfy 
the graded reflection equations \cite{Bra98}:
\bea
&&R_{12}(u_1-u_2)\stackrel {1} {K}_-(u_1) R_{21}(u_1+u_2)
\stackrel {2}{K}_-(u_2)= 
\stackrel {2}{K}_-(u_2) R_{12}(u_1+u_2)
\stackrel {1}{K}_-(u_1) R_{21}(u_1-u_2),\no\\ 
&&R_{12}(-u_1+u_2)\stackrel {1}{K_+}(u_1)\stackrel{1}{M^{-1}} 
R_{21}(-u_1-u_2)\stackrel{1}{M}
\stackrel {2}{K_+}(u_2)\no\\
&&~~~~~~~~~~~~~~~~~~~~~~~~~~ =\stackrel{1}{M} 
\stackrel {2}{K_+}(u_2) R_{12}(-u_1-u_2)\stackrel{1}{M^{-1}}
\stackrel {1}{K_+}(u_1) R_{21}(-u_1+u_2),\label{RE-with-cu}
\eea
where $\stackrel {1}{X} \equiv X \otimes 1$ and 
$\stackrel {2}{X} \equiv 1 \otimes X$ , for any matrix $ X \in End(V) $.
$R(u)$ is the quantum $R$-matrix of the
$q$-deformed EKS model:  $R(u) \equiv
P \check {R}(u)$, with $\check {R}(u)$ as in (\ref{rational-R}),
and $R_{21}(u)= P_{12}R_{12}(u)P_{12}$ with $P$ being the graded permutation 
operator. 

In order to describe integrable systems with boundary conditions different
from periodic ones, we first solve the reflection equations for the two
boundary $K$-matrices $K_{\pm}(u)$. 
For our purpose, we only look for solutions where $K_\pm(u)$ are
diagonal. After complicated algebraic manipulations, we find three 
different diagonal boundary $K$-matrices, $K^I_-(u),~
K^{II}_-(u),~K^{III}_-(u)$, which solve the first reflection equation in
(\ref{RE-with-cu}):
\beq
K^I_-(u)=  \frac {1}{ (\sinh \frac {\g \xi^I_-}{2})^2}\;
\left ( \begin {array}{cccc}
A^I_-(u)&0&0&0\\
0&B^I_-(u)&0&0\\
0&0&C^I_-(u)&0\\
0&0&0&C^I_-(u)
\end {array} \right ),\label{k^I-}
\eeq
\beq
K^{II}_-(u)=  \frac {1}{ (\sinh \frac {\g \xi^{II}_-}{2})^2}\;
\left ( \begin {array}{cccc}
A^{II}_-(u)&0&0&0\\
0&B^{II}_-(u)&0&0\\
0&0&B^{II}_-(u)&0\\
0&0&0&C^{II}_-(u)
\end {array} \right ),\label{k^II-}
\eeq
and
\beq
K^{III}_-(u)=  \frac {1}{ \sinh \frac {\g \xi^{III}_-}{2}}\;
\left ( \begin {array}{cccc}
A^{III}_-(u)&0&0&0\\
0&A^{III}_-(u)&0&0\\
0&0&B^{III}_-(u)&0\\
0&0&0&B^{III}_-(u)
\end {array} \right ),\label{k^III-}
\eeq
where
\bea
A^I_-(u)&=&e^{-\g u} \sinh  \frac {\g (\xi^I_-+u)}{2}\sinh \frac {\g (-u+\xi^I_-)}{2},\no\\
B^I_-(u)&=&( \sinh \frac {\g (\xi^I_-+u)}{2})^2,\no\\
C^I_-(u)&=&e^{\g u} \sinh \frac {\g (\xi^I_-+u)}{2}\sinh \frac {\g (-u+\xi^I_-)}{2},
\no\\
A^{II}_-(u)&=&e^{-\g u} \sinh  \frac {\g (\xi^{II}_-+u)}{2}\sinh \frac {\g (-u+
\xi^{II}_-)}{2},\no\\
B^{II}_-(u)&=&( \sinh \frac {\g (\xi^{II}_-+u)}{2})^2,\no\\
C^{II}_-(u)&=&e^{\g u} \sinh \frac {\g (\xi^{II}_-+u)}{2}\sinh \frac {\g (-u+
\xi^{II}_-)}{2},\no\\
A^{III}_-(u)&=&e^{-\frac{\g u}{2}} \sinh \frac {\g (-u+ \xi^{III}_-)}{2},\no\\
B^{III}_-(u)&=&e^{\frac{\g u}{2}} \sinh \frac {\g (u+\xi^{III}_-)}{2}.
\eea
As is shown in \cite{Bra98,Lin96}, the $R$-matrix (\ref{rational-R}) satisfies
the crossing-unitarity condition.  
The corresponding $K$-matrices $K^I_+(u),~K^{II}_+(u)$ and $K^{III}_+(u)$ which
obey the second reflection
equation in (\ref{RE-with-cu}) may be derived by isomorphism:
\beq
K^I_+(u)=M K^I_-(-u),~~~~~~K^{II}_+(u)=MK^{II}_-(-u)
,~~~~~~K^{III}_+(u)=MK^{III}_-(-u)
\eeq
where $M$ is the so-called crossing matrix given by: 
\beq
M=\left ( \begin {array} {cccc}
1&0&0&0\\
0&e^{2\g}&0&0\\
0&0& e^{2\g}&0\\
0&0&0&1 
\end {array} \right ).\label{m}
\eeq
Therefore we may choose the boundary $K$-matrices $K^{I}_+(u),K^{II}_+(u) $
and $K^{III}_+(u)$ as
\beq
K^I_+(u)=\left ( \begin {array} {cccc}
A^I_+(u)&0&0&0\\
0&B^I_+(u)&0&0\\
0&0& C^I_+(u)&0\\
0&0&0& D^I_+(u)
\end {array} \right )\label{k^I+}
\eeq
\beq
K^{II}_+(u)=\left ( \begin {array} {cccc}
A^{II}_+(u)&0&0&0\\
0&B^{II}_+(u)&0&0\\
0&0&B^{II}_+(u)&0\\
0&0&0&C^{II}_+(u)
\end {array} \right )\label{k^II+}
\eeq
and
\beq
K^{III}_+(u)=\left ( \begin {array} {cccc}
A^{III}_+(u)&0&0&0\\
0&B^{III}_+(u)&0&0\\
0&0&C^{III}_+(u)&0\\
0&0&0&D^{III}_+(u)
\end {array} \right )\label{k^III+}
\eeq
with
\bea
A^I_+(u)&=& ( \sinh \frac {\g(u-2+\xi^I_+)}{2})^2,\no\\
B^I_+(u)&=&e^{-2\g}( \sinh \frac {\g(u+2-\xi^I_+)}{2})^2,\no\\
C^I_+(u)&=&e^{(-2-u)\g} \sinh \frac {\g(u-2+\xi^I_+)}{2}
\sinh \frac {\g(-u-2+\xi^I_+)}{2},\no\\
D^I_+(u)&=&e^{-u\g}\sinh \frac {\g(u-2+\xi^I_+)}{2}
\sinh \frac {\g(-u-2+\xi^I_+)}{2},\no\\ 
A^{II}_+(u)&=&e^{\g u} \sinh \frac {\g(u+\xi^{II}_+)}{2}
\sinh \frac {\g(-u+\xi^{II}_+)}{2},\no\\
B^{II}_+(u)&=&e^{-2\g}( \sinh \frac {\g(u-\xi^{II}_+)}{2})^2,\no\\
C^{II}_+(u)&=&e^{(-u)\g} \sinh \frac {\g(u+\xi^{II}+)}{2}
\sinh \frac {\g(-u+\xi^{II}_+)}{2},\no\\
A^{III}_+(u)&=&e^{\frac{\g u}{2}} \sinh \frac {\g(u+4+\xi^{III}_+)}{2},\no\\
B^{III}_+(u)&=&e^{\frac{\g(u-4)}{2}}\sinh \frac {\g(u+4-\xi^{III}_+)}{2},\no\\
C^{III}_+(u)&=&e^{-\frac{\g(u+4)}{2}} \sinh \frac {\g(-u+4+\xi^{III}_+)}{2} ,\no\\
D^{III}_+(u)&=&e^{-\frac{\g u}{2}}\sinh \frac {\g(-u+4+\xi^I_+)}{2}.
\eea
It can be checked explicitly that these $K_+$ matrices constitute solutions
to the second reflection equation in (\ref{RE-with-cu}).
 
We form the boundary transfer matrix $t(u)$:
\beq
t(u)=str[K_+(u)T_-(u)K_-(u)T^{-1}_-(-u)],~~~~~~
  T_-(u) = R_{0L}(u) \cdots R_{01}(u).
\eeq
Since $K_\pm(u)$  can be taken as $K^I_\pm(u)$ or $K^{II}_\pm(u)$ or
$K^{III}_\pm(u)$,
respectively, we have nine possible choices of the boundary transfer matrices:
\bea
t^i(u)&=&str[K^I_+(u)T_-(u)K^I_-(u)T^{-1}_-(-u)],\no\\
t^{ii}(u)&=&str[K^{II}_+(u)T_-(u)K^{II}_-(u)T^{-1}_-(-u)],\no\\
t^{iii}(u)&=&str[K^{III}_+(u)T_-(u)K^{III}_-(u)T^{-1}_-(-u)],\no\\
t^{iv}(u)&=&str[K^{II}_+(u)T_-(u)K^I_-(u)T^{-1}_-(-u)],\no\\
t^{v}(u)&=&str[K^I_+(u)T_-(u)K^{II}_-(u)T^{-1}_-(-u)],\no\\
t^{vi}(u)&=&str[K^{III}_+(u)T_-(u)K^I_-(u)T^{-1}_-(-u)],\no\\
t^{vii}(u)&=&str[K^I_+(u)T_-(u)K^{III}_-(u)T^{-1}_-(-u)],\no\\
t^{viii}(u)&=&str[K^{III}_+(u)T_-(u)K^{II}_-(u)T^{-1}_-(-u)],\no\\
t^{ix}(u)&=&str[K^{II}_+(u)T_-(u)K^{III}_-(u)T^{-1}_-(-u)].\label{t-matrices}
\eea
which reflects the fact that 
the boundary conditions on the left end and on the right end of
the open lattice chain are independent.
Substituting these expressions into the boundary 
transfer matrix $t(u)$, and after a
lengthy but straightforward algebraic calculation, one finds
\beq
t(u) =  C_1 u + C_2 (H + const.) u^2 + \cdot\cdot\cdot,
\eeq
where $C_i (i = 1,2,\cdots)$ are some scalar functions of the boundary
constant $\xi _+$. Then it can be shown that up to some additive 
constant,  the boundary Hamiltonians with the boundary conditions 
(\ref{boundary11}--{\ref{boundary32}) are related to
the second derivative of the corresponding boundary transfer matrix
\cite{Zha97}: 
\bea
H&=&-\frac{q-q^{-1}}{\ln q}H^R,\no\\
H^R&=&\frac {t'' (0)}{4(V+2W)}=
  \sum _{j=1}^{L-1} H^R_{j,j+1} + \frac {1}{2} \stackrel {1}{K'}_-(0)
+\frac {1}{2(V+2W)}\lt[str_0\lt(\stackrel {0}{K}_+(0)G_{L0}\rt)\rt.\no\\
& &\lt.+2\,str_0\lt(\stackrel {0}{K'}_+(0)H_{L0}^R\rt)+
  str_0\lt(\stackrel {0}{K}_+(0)\lt(H^R_{L0}\rt)^2\rt)\rt],\label{derived-h}
\eea
where  
\bea
V&=&str_0 K'_+(0),
~W=str_0 \lt(\stackrel {0}{K}_+(0) H_{L0}^R\rt),\no\\
H^R_{i,j}&=&P_{i,j}R'_{i,j}(0),
~G_{i,j}=P_{i,j}R''_{i,j}(0).
\eea
If we make the identification $q=e^{\g}$,
we then see that the  Hamiltonians (\ref{hamiltonian-b}) of the 
$q$-deformed EKS model corresponding to all nine boundary
conditions can be determined from the above nine boundary transfer
matrices, respectively. We thus arrive at the nine possible
cases (\ref{boundary11}--{\ref{boundary32}), all of which are
compatible with the bulk integrability.

\vskip.3in
The author is supported by OPRS and UQPRS.
He would like to express his sincere thanks to
A.J.Bracken for his encouragement.
His appreciation also goes to Y.-Z. Zhang and
H.-Q. Zhou for suggesting and useful dicussions.

\end{document}